# Navigating Munk's Abyssal Recipes: Reconciling the Paradoxes and Suggesting an Upwelling Mechanism for Bottom Water in a Flat-Bottom Ocean

Lei Han [a, b]

[a] *China-ASEAN College of Marine Sciences, Xiamen University Malaysia, Sepang, Malaysia*

[b] *College of Ocean and Earth Sciences, Xiamen University, Xiamen, China*

*Corresponding author*: Lei Han, lei.han@xmu.edu.my

[Supplementary materials are appended at the end of this document]

# ABSTRACT

Walter Munk's classical work, known as the "abyssal recipes", introduced a foundational framework for comprehending the upwelling of abyssal waters. While it has spurred numerous investigations into the complexities of deep-ocean processes from theoretical, laboratory, and field perspectives, it has also faced challenges when compared to direct observations spanning decades. Two particularly intriguing paradoxes emerge: the dichotomy of diffusivities and the conundrum of interior downwelling. This study attempts to resolve these paradoxes by examining the isopycnal displacement velocity within the dynamic framework of Munk's abyssal recipes. A relieving inference is that it seems no longer an imperative to seek a globally averaged diapycnal diffusivity of $O(1)$ cm$^2$/s to close the mass budget associated with the bottom-water formation rate. Through box-model experiments, a novel erosion-intrusion model is proposed to elucidate the mechanism of abyssal upwelling in flat-bottom ocean, drawing an analogy to the growth of "tree rings". Based upon this model, the average rising velocity in the abyssal North Pacific is estimated using observational data of biogeochemical tracers. The proposed model does not seek to disprove existing theories but rather acts as a complement for situations that prevailing theories may not apply, such as cases with flat-bottom topography or non-bottom-intensified diffusivity.

# SIGNIFICANCE STATEMENT

This study revisits an important theory in ocean dynamics from 1966 that deals with how deep ocean waters move upward to the surface, and attempts to unravel long-standing mysteries related to this theory. Additionally, I propose an innovative model to explain how the water rises from the flat-bottom seafloor based on a series of idealized numerical experiments. This model offers a new explanation on cases of abyssal upwelling that the existing theories do not apply. It draws an analogy between the abyssal upwelling and growth of tree rings. This work matters because it provides fresh insights into the long unclear question of how the coldest water returns from the bottom of the ocean to close the global conveyor belt.

# 1. Introduction

The global meridional overturning circulation, often referred to as the global conveyor belt, plays a pivotal role in climate by wielding immense power in transporting heat. It comprises two key limbs: the downwelling limb, where dense water formed in the high-latitude ocean descends to great depths, and the upwelling limb, responsible for bringing deep waters back to the surface across the remaining ocean basins. While the downward process of "push", propelled by deep-water formation, is well-understood, the upward pathway of "pull" has long been a missing "crucial piece of the giant marine puzzle" (e.g., Visbeck 2007; Marshall and Speer 2012; Ferrari 2014; Voosen 2022).

One leg of the return path has become clearer in recent years through the tilted isopycnal layers that ventilate or outcrop in the Southern Ocean and extend downwards into the intermediate depths to the north (Toggweiler and Samuels 1998; Marshall and Speer 2012; Ferrari 2014; Rintoul 2018). The return journey along this path is cost-free for the water parcels above the depth of 3500 meters that seek to rise back to the surface, as no additional energy or water transformation is needed when traveling along an isopycnal (Tamsitt et al 2017), although the diapycnal processes may contribute a minor proportion (Wolfe and Cessi 2011; Watson et al 2013; Naveira Garabato et al 2016). However, the homeward voyage for water parcels below 3500 meters is not as straightforward. They must decrease in density before they can ascend into this "open tunnel" leading back to the Southern Ocean. This part of upwelling, the intensity of which cannot be adequately accounted for by geothermal heating alone (e.g., Emile-Geay and Madec 2008; de Lavergne et al 2016), stands as the most enigmatic aspect of the entire return pathway.

Efforts to solve this abyssal rising pathway date back to Walter Munk's groundbreaking paper titled "Abyssal recipes" (Munk 1966), which has garnered around 1700 citations to date (Wunsch 2023). By assuming a steady background, Munk proposed a simple 1D vertical advection-diffusion balance of the form,

$$w\frac{\partial \rho}{\partial z} = \kappa \frac{\partial^2 \rho}{\partial z^2}, \tag{1}$$

with $w$ the Eulerian vertical velocity and $\kappa$ the uniform vertical diffusivity. Fitting the density and radiocarbon profiles from the central Pacific led to values of $w=1.4\times10^{-7}$ m/s and $\kappa=1.3\times10^{-4}$ m$^2$/s. As supporting evidence, it was argued that assuming spatially uniform upwelling, the value of $w$ was broadly consistent with the bottom-water formation rate at high latitudes. Inferred from mass-budget closure, subsequent observational estimates affirmed $w$ to be on the order of $10^{-7}$ m/s at depth in the global ocean (e.g., Hogg et al 1982; Stuiver et al 1983; Wunsch et al 1983; Warren and Speer 1991; Speer and Zenk 1993; Robbins and Bryden 1994; Robbins and Toole 1997). Three decades later, the paper "Abyssal recipes II" updated the 1D advection-diffusion model by incorporating horizontal advections, although the estimated $\kappa$ remained on the order of $10^{-4}$ m$^2$/s to maintain the abyssal stratification (Munk and Wunsch 1998).

In contrast, the microstructure measurements and dye release experiments both found that the average vertical diffusivity in the ocean was around $1\times10^{-5}$ m$^2$/s, which is one order of magnitude smaller than what was predicted by the abyssal recipes (e.g., Osborn and Cox 1972; Stigebrandt 1979; Gregg 1989; Ledwell et al 1993; Toole et al 1994; Kunze and Sanford 1996; Kunze et al 2006). This discrepancy, known as the "dichotomy of diffusivities" (Munk and Wunsch 1998), has been consistently observed in subsequent measurements of $\kappa$ (e.g., Kunze 2017a,b; Goto et al 2021). Additionally, it has been found that $\kappa$ has a Gaussian-like probability distribution, with peaks around $1\times10^{-5}$ m$^2$/s rather than $1\times10^{-4}$ m$^2$/s (Kunze and Sanford 1996; Martin and Rudnick 2007; Goto et al 2021). An extensive compilation of microstructure measurements also demonstrates that the majority of available measurements indicates a distribution center to the left of $1\times10^{-4}$ m$^2$/s at most depths (see Fig. 6b in Waterhouse et al 2014).

Efforts to uncover the "missing mixing" have been partial success through the discovery of significantly elevated diffusivities near the seafloor, particularly in areas characterized by topographic features and regions with heightened internal-wave energy (e.g., Toole et al 1994; Polzin et al 1997; Kunze et al 2006; Lee et al 2006; Waterman et al 2013; Waterhouse et al 2014; Voet et al 2015; Mashayek et al 2017; Kunze 2017a,b). Initially aimed at reconciling Munk's theory, these observations have unexpectedly opened *Pandora's box* by introducing a more profound challenge to the theory: The bottom intensification of the turbulent dissipation rate has been recognized as a consistent feature in various field campaigns (e.g., Polzin et al 1996; Morris et al 2001; St. Laurent et al 2001; Waterman et al 2013; Mashayek et al 2017). As extensively reviewed in numerous studies, the bottom-intensified dissipation rate, denoted as $\varepsilon$, suggests a downwelling (*w*<0) rather than an upwelling (*w*>0) according to Munk's theory (e.g., St. Laurent et al 2001; Ferrari et al 2016; Callies and Ferrari 2018; Drake et al 2020). This is because its diffusion term, equivalent to the divergence of the turbulent density flux, can be related with $\varepsilon$ as follows (St. Laurent et al 2001),

$$w = N^{-2}\frac{\partial}{\partial z}(\Gamma\varepsilon), \qquad (2)$$

where $N^2$ is the buoyancy gradient and $\Gamma$ the mixing efficiency (Osborn 1980; Gregg et al 2018). The inference of downwelling contradicts the mass balance in the deep ocean, posing a "conundrum" that is even more destructive to Munk's theory than the dichotomy of diffusivities.

Motivated by this conundrum, subsequent theoretical studies have shifted their focus to an alternative perspective. They argue that an upwelling is driven by highly localized turbulence within thin layers near the seafloor, typically tens of meters in thickness. Building upon this concept, models have been developed that include a downwelling interior layer and an upwelling bottom boundary layer (BBL) (e.g., de Lavergne et al 2016; Ferrari et al 2016; de Lavergne et al 2017; McDougall and Ferrari 2017; Callies and Ferrari 2018; Drake et al 2020; Ruan and Callies 2020; Drake et al 2022; Drake et al 2022). According to estimates by McDougall and Ferrari (2017), to achieve a net transport of 18 Sv through upwelling, the upwelling BBL would need to support a vertical volume transport of ~108 Sv. This level of transport within the BBL, spanning approximately 0.2° of longitude or latitude with a layer thickness of around 50 m, corresponds to a vertical velocity on the order of $O(10^{-3})$ m/s. This velocity is four orders of magnitude greater than that of uniform upwelling. The turbulent diffusivity immediately above the BBL would need to be ~$5\times10^{-3}$ m$^2$/s.

A preliminary finding from an ongoing dedicated observational project in the Rockall Trough revealed an upward velocity of 100 m per day or $1.2\times10^{-3}$ m/s (Voosen 2022). However, another observational study conducted near the foot of a steep East-Pacific continental slope found no evidence of "turbulent bores propagating up the slope", even with a high-resolution instrumentation sampling from within 0.5 m to 150 m above the seafloor (van Haren et al 2022). Additionally, in an experiment in which a turbulent boundary layer was generated along a sloping wall of a laboratory tank containing salt-stratified fluid, researchers observed not only upslope flow but also downslope flow within the BBL (Phillips et al 1986).

Distinct from the upwelling BBL theories, this study attempts to reconcile the conundrum of Munk's theory from another perspective. Steady-state was assumed in the vertical balance by arguing that "there were no significant alterations in 21 years" (Munk 1966) and also adopted in some extended works (e.g., Munk and Wunsch 1998; Wunsch 2023). A recent study, comparing data collected from the HMS Challenger expedition of the 1870s and modern hydrography, suggests that the deep Pacific Ocean has cooled by 0.02 °C over the past century (Gebbie and Huybers 2019). This corresponds to a change of ~0.004 °C in a 20-year period. Reviewing global temperature observations, it is noted that temperature measurement accuracy in the 1960s and earlier was typically no better than 0.1 °C (Abraham et al 2013). Hence, it would have been impossible to detect such a minuscule temperature difference with the techniques available during that time.

Nevertheless, this minute change is crucial to the vertical balance in the abyssal recipes. Repeated occupations of hydrographic sections in the Pacific basin have noted consistent deep ocean changes, "dispelling the notion that the deep ocean is quiescent" (Sloyan et al 2013). Volume budget below deep, cold isotherm within the Pacific basin "are not in steady state" (Purkey and Johnson 2012). Field observations reveal that the displacement trend of isopycnal/isotherm in the bottom ocean reaches the order of several meters per year, comparable to the averaged Eulerian upwelling velocity balancing the bottom-water formation (e.g., Purkey and Johnson 2012; Sloyan et al 2013; Voet et al 2016; Purkey et al 2019; Zhou et al 2023). These observational facts necessitate the serious consideration of the non-stationarity in the advection-diffusion balance. Based on this recognition, this study demonstrates that the unsteady 1D advection-diffusion balance does have the potential to reconcile the conundrum of Munk's theory, as well as address the missing mixing that has long perplexed field investigators.

Regarding the mechanism of abyssal upwelling, instances have arisen that fall outside the applicability of the upwelling BBL theory, such as cases of flat-bottom topography or non-bottom-intensified mixing. This study conducts a series of numerical experiments using a box model to investigate these occasions. Based on the experimental outcomes, an innovative conceptual model concerning erosion-intrusion process is proposed. The new model does not aim to refute the upwelling BBL theory; instead, it serves as a supplement for situations where the BBL theory is not applicable. Finally, the biogeochemical tracer data from a comprehensive dataset are utilized to estimate the upwelling velocity in the deep Pacific Ocean, providing a spatial distribution by stations.

This paper is organized as follows. Section 2 explains how the conundrum of interior downwelling is reconciled by considering the displacement velocity of isopycnals in the unsteady advection-diffusion balance. Section 3 presents the box-model experiments, on which an erosion-intrusion model accounting for the Eulerian upwelling of the bottom water in a flat-bottom ocean is proposed. Section 4 presents an estimation of the averaged upwelling velocity across the deep Pacific Ocean using tracer data. Finally, Section 5 offers a summary and discussions.

## 2. Reconciling the conundrum of Munk's theory

Though Munk and his collaborator presumed steady balance (Munk 1966; Munk and Wunsch 1998; Wunsch 2023), unsteady abyssal recipes have been derived and are now widely used in a modified form as follows (for density) (e.g., McDougall 1984; Marshall et al 1999; St. Laurent et al 2001; de Lavergne et al 2016; Ferrari et al 2016),

$$(w_{Eul} - w_{iso})\frac{\partial \rho}{\partial z} = -\frac{\rho}{g}\frac{\partial}{\partial z}(\Gamma \varepsilon), \qquad (3)$$

with various expressions of the diffusion term (RHS) as follows,

$$-\frac{\rho}{g}\frac{\partial}{\partial z}(\Gamma\varepsilon) = \frac{\partial}{\partial z}\left(\kappa\frac{\partial\rho}{\partial z}\right) = -\frac{\partial F_\rho^z}{\partial z} = \frac{\partial F_b^z}{\partial z}, \tag{4}$$

where $F_\rho^z = \overline{w'\rho'}$ and $F_b^z = \overline{w'b'}$ represent the vertical turbulent density flux and buoyancy flux, respectively. Eq (3) can be simply written as,

$$w_{dia}\frac{\partial\rho}{\partial z} = -\frac{\rho}{g}\frac{\partial}{\partial z}(\Gamma\varepsilon), \tag{5}$$

with $w_{dia} = w_{Eul} - w_{iso}$. Here, the three "$w$" variables denote the Eulerian vertical velocity ($w_{Eul}$), the isopycnal vertical velocity ($w_{iso}$), and the diapycnal vertical velocity ($w_{dia}$), respectively. $w_{iso}$ stems from the time derivative of density and represents the unsteady effect (Marshall et al 1999). Connections among the three velocities were illustrated in Fig. 4 of Han (2021). Above equations can be also converted to a vector form (Marshall et al 1999; Ferrari et al 2016), but the vertical component suffices for illustrating the main point in the subsequent analysis. The concept of its vector form will be utilized in section 4.

While the unsteady abyssal recipes are widely adopted in many recent studies, the change of the background stratification, specifically isopycnal displacement, $w_{iso}$, is unfortunately omitted. In such cases, the unsteady recipe, as expressed in Eq. (5), reduces to the steady case, Eq. (1) or (2). How critical is the isopycnal vertical velocity, $w_{iso}$, to the mass budget closure in the abyssal circulation? As a first step, let's examine the density variation in time in the deep ocean.

**2.1 A cooling abyss in the past decades**

Here I examine the density change of the deep ocean with the Estimating the Circulation and Climate of the Ocean State Estimate version 4 release 3 (ECCO v4r3) (Forget et al 2015; Fukumori et al 2017). This product is generated by best fitting the numerical simulation of an ocean general circulation model, the Massachusetts Institute of Technology general circulation model (MITgcm) to over 1 billion observations (Rousselet et al 2020; Rousselet et al 2022). The ECCO v4r3 utilized in this study has a nominal horizontal resolution of 1° and a depth coordinate system with 50 levels. It covers the time period from 1992 to 2015 with a monthly resolution. This dataset has been employed in a recent study that also revisited Munk's abyssal recipes (Wunsch

2023).

The potential density referenced to 4 km ($\sigma_4$) at a depth of 4264 m is chosen for illustration. It shows the difference between two 5-year periods, 2011−2015 and 1995−1999 (Fig. 1). The chosen averaging period is to minimize the impact of seasonal variability as previously noted in the abyssal Indian Ocean (Han 2021). Fig. 1 reveals a noteworthy spatial distribution with opposite trends. Though the warming and freshening trends in the Antarctic Bottom Water (AABW) have been consistently documented in repeat hydrographic sections(Rintoul 2007; Zenk and Morozov 2007; Menezes et al 2017; Purkey et al 2019; Zhou et al 2023), the majority of the abyssal basins shown in Fig. 1a is experiencing densification. The principal driver of this densification is the cooling temperature (Fig. 1b, c). Other studies have also identified a cooling abyss in ECCO (Wunsch and Heimbach 2014; Liang et al 2015).

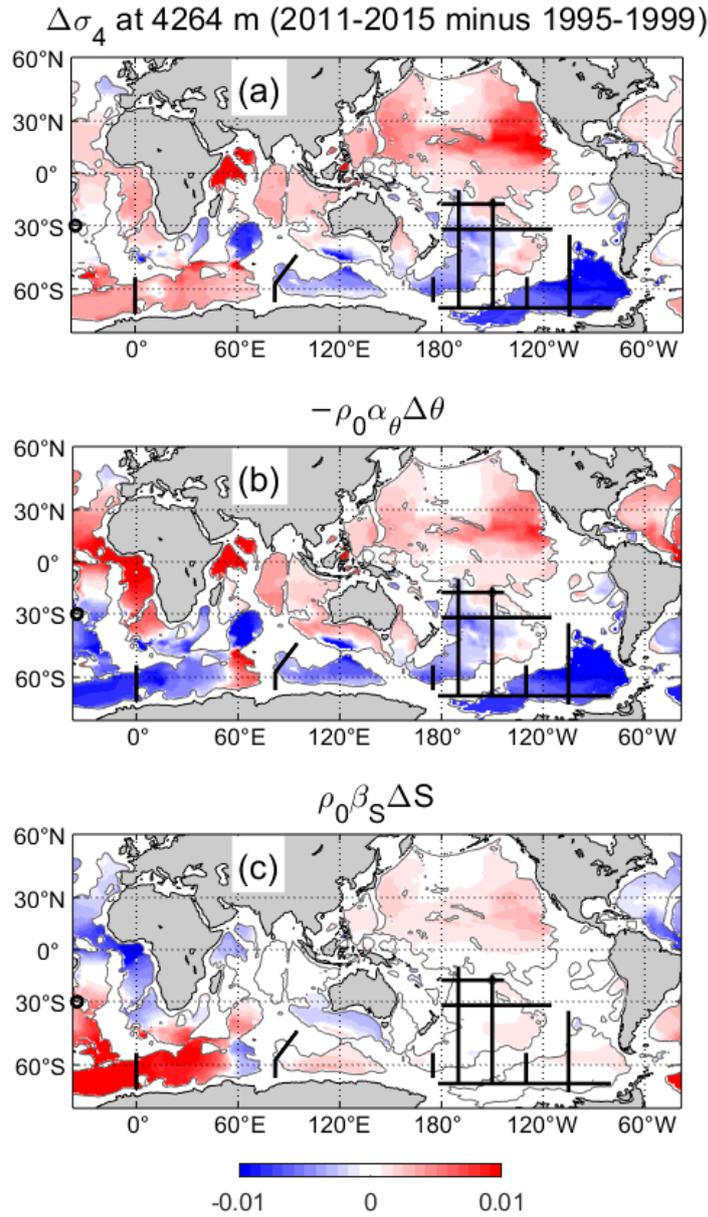

Fig. 1. The change of potential density (unit: kg/m$^3$) referenced to 4km ($\sigma_4$) at a depth of 4264 m between two 5-year-period averages (2011-2015 minus 1995-1999) (a). The relative contributions from potential temperature (b) and salinity (c) to this density change are estimated by virtue of the local thermal expansion coefficient, $\alpha_\theta$, and the saline contraction coefficient, $\beta_S$. Black solid lines/circle in each panel mark the repeat hydrographic sections/station that have revealed a warming trend of AABW around the Antarctic (Pacific sector: Purkey et al 2019; Indian Ocean sector: Menezes et al 2017; Atlantic sector: Zhou et al 2023; Vema Channel: Zenk and Morozov 2007). The warm color represents densification(a), cooling(b), or salinification(c) while the cold color represents the opposite changes. Data: ECCO v4r3.

Given the backdrop of global warming, is the cooling trend in the abyssal basins a genuine phenomenon or a data error? Can we trust the ECCO product for exploring the mechanism of the abyssal circulations? How to reconcile the surface warming and abyssal cooling? A recent study has provided important insights into these puzzles. Combining an ocean model with modern and palaeoceanographic data, Gebbie and Huybers (2019) argued that the deep Pacific is still adjusting to the cooling going into the Little Ice Age from 1400 to 1900 AD. This cooling trend is corroborated by temperature changes identified between the HMS Challenger expedition of the 1870s and modern hydrography. Their model found the deep Pacific has cooled by 0.02 °C over the past century. The time series of the surface temperature showed the cooling course in the Antarctic region began around 600 AD and peaked in 1400−1600 AD. According to the water-age distribution in the deep ocean (>2 km) estimated by a data-constrained ocean circulation model (Fig. 12a in DeVries and Primeau 2011), the age of the oldest water in the North Pacific is around 1400 years. This suggests that the first cooling signal associated with the Little Ice Age in the Antarctic region may have just spread throughout the entire Pacific. Furthermore, the first warming signal, which commenced after the peak of the Little Ice Age in around 1400−1600 AD, has spread for 400−600 years so far. As per the water-age distribution in the deep ocean estimated with radiocarbon observations (e.g., Sarmiento and Gruber 2006; Matsumoto 2007; DeVries and Primeau 2011; Gebbie and Huybers 2012), this warming signal should have just swept through the Southern Ocean area as it spreads northwards. This is in broad agreement with the ECCO result (Fig. 1b). The movie of the 2000-year evolution in Gebbie and Huybers (2019) clearly illustrates the northward spread of the potential temperature anomaly at a depth of 2500 m from the Southern Ocean into the three tropical oceans.

Additionally, the abyssal trends from ECCO are in broad agreement with the observed trends revealed by the repeat hydrographic sections. Fig. 1 illustrates locations of the sections where a warming and freshening AABW was recognized. With a similar comparison period as the repeat hydrographic sections, ECCO data also shows warming along those sections (Fig. 1b). However, the freshening trend is not profound in ECCO for those sections. More critically, there is an excessively large salinification trend in the Atlantic sector (Fig. 1c). This increase in salinity is so substantial that it outweighs the effect of warming, leading to densification. This result contradicts the repeat hydrographic data (Zhou et al 2023).

To summarize, ECCO skillfully reproduces the temperature trends in the deep ocean during its time span but exhibits limitations in simulating salinity changes. Given that temperature dominates the density change in most regions, particularly the Indo-Pacific region, where the abyssal upwelling primarily occurs, ECCO is deemed acceptable for analyzing bottom-water upwelling in the following sections.

**2.2 Diapycnal downwelling in the ocean interior**

The streamfunctions of meridional overturning circulation (MOC) are commonly used to visualize the volume transport in the vertical plane. Here four types of MOC streamfunctions for the Indo-Pacific Ocean are demonstrated. By definition, the MOC streamfunction can be derived with either the meridional velocity component (*v*) or the vertical velocity component (*w*) for a basin with closed zonal boundary such as the Indo-Pacific basin to the north of 34°S. Stemming from this concept, I used to define two new MOC streamfunctions using vertical velocity $w_{iso}$ and $w_{dia}$ for studying the overturning dynamics in the Indian Ocean and the Atlantic Ocean, respectively (Han 2021,2023). The first is the sloshing MOC streamfunction, defined as $\psi_{iso}$, which characterizes the overturning circulations associated with vertical displacement of isopycnals. The second is the diapycnal MOC streamfunction, denoted as $\psi_{dia}$, which is defined as the difference between the conventional Eulerian MOC streamfunction (denoted as $\psi_{Eul}$) and $\psi_{iso}$. It represents the diabatic component of the overturning circulations. For example, in the case of adiabatic process, the material surface and isopycnal remain attached to each other throughout. In these instances, $\psi_{iso}$ is equal to $\psi_{Eul}$, resulting in $\psi_{dia}$ being equal to zero. As $\psi_{Eul}$ and $\psi_{iso}$ diverge from each other (i.e., diabatic process), $\psi_{dia}$ increases in magnitude, signifying an increasing role of the diabatic processes.

The three full-time-averaged MOC streamfunctions for the Indo-Pacific Ocean are computed with ECCO (Fig. 2). To affirm the upwelled water is indeed transformed (unlike the Deacon cell in the Southern Ocean [e.g., Döös et al 2008]), I also compute the streamfunction in the density coordinate of $\sigma_4$ (denoted as $\psi_\sigma$).

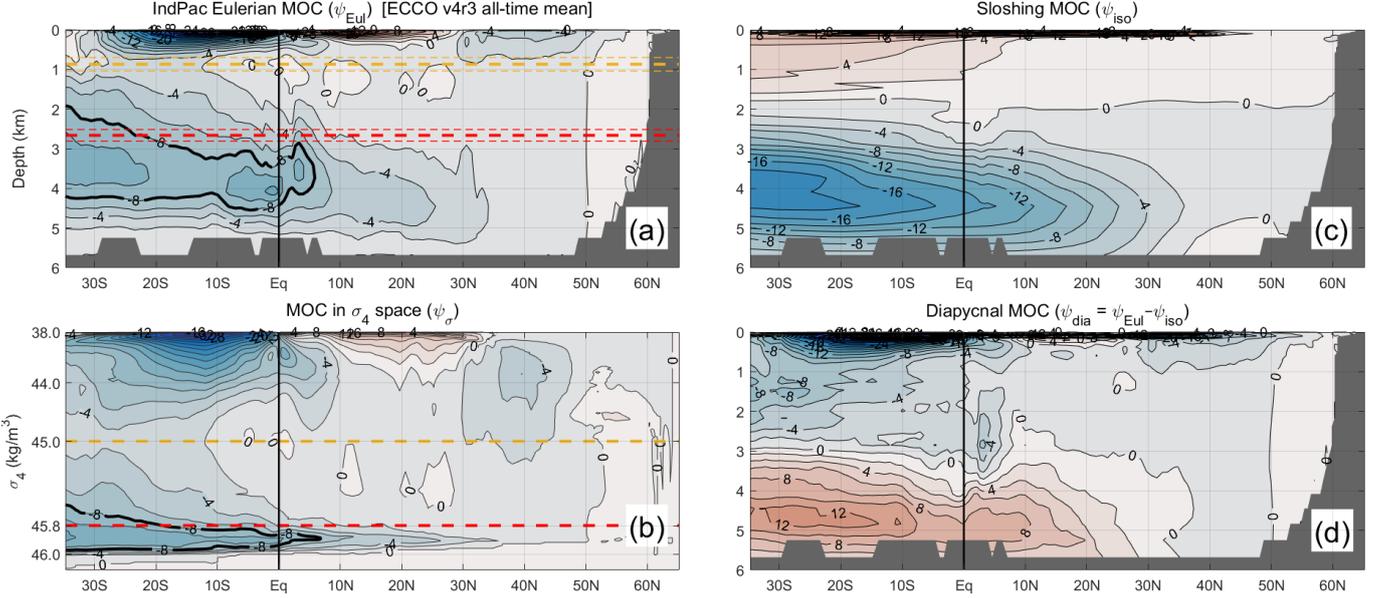

Fig. 2. The full-time-averaged streamfunctions of overturning in the Indo-Pacific Ocean. (a) Eulerian MOC streamfunction; (b) Streamfunction in the density coordinate of $\sigma_4$; (c) Sloshing MOC streamfunction; (d) Diapycnal MOC streamfunction. Positive values (warm color) denote clockwise-overturning cells while negative values (cold color) denote anti-clockwise (upwelling) cells. The thick (thin) dashed lines in panel (a) represent the mean (s.d.) depth of the two isopycnals, $45.0\sigma_4$ and $45.8\sigma_4$, respectively, as indicated in panel (b). The contour of -8 Sv in the deep cell is thickened for clearer comparison in (a) and (b). Data: ECCO v4r3.

As shown in Fig. 2a, a large-scale anti-clockwise deep overturning cell is remarkable in $\psi_{Eul}$, indicating a significant upwelling of the bottom water to the intermediate depths. Other studies using the same dataset showed a similar pattern of this cell though they showed the residual MOC streamfunctions (Rousselet et al 2021; Rogers et al 2023). Difference between the Eulerian and residual MOC streamfunction is discussed in details in the following section. When comparing streamfunctions in depth and density coordinates, it becomes evident that the deep cells in the two solutions exhibit a close agreement in terms of extent and magnitude. This suggests that the upwelled water in the Eulerian coordinate has undergone significant transformation into lighter densities in the Indo-Pacific Ocean before exiting the basin.

Now if we look at the isopycnal change that is illustrated by the sloshing MOC, $\psi_{iso}$ (Fig. 2c), it is surprising to find the abyssal isopycnals are moving upward at a speed even faster than the water parcels, which is described by $\psi_{Eul}$ (Fig. 2a)! Given that temperature dominates the density change in the deep ocean, this phenomenon corresponds to an abyssal cooling, which is consistent with the observed vertical divergence of turbulent heat flux associated with bottom-intensified mixing. Meanwhile, this difference between $w_{Eul}$ and $w_{iso}$ leads to a negative diapycnal vertical velocity, $w_{dia}$, which is shown as a remarkable downwelling overturning cell in Fig. 2d. According to the unsteady abyssal recipes, Eq. (5), this explains that the *bottom-intensified dissipation indeed produces a downward velocity*, only that this vertical velocity is not the Eulerian one, but the diapycnal one. Moreover, this diapycnal downwelling has nothing to do with the physical movement of water parcels; it occurs simply due to the presence of cooling (Fig. 4b of Han 2021). In other words, the interior of the abyssal ocean is still experiencing upwelling, only that the water is simultaneously cooled by the bottom-intensified dissipation (or equivalently, bottom-intensified density flux, according to Eq. (4)) by losing more heat to the layer below than that gained from the layer above as it rises. Therefore, the conundrum can be resolved if the downward diapycnal vertical velocity is properly interpreted. In this sense, the bottom-intensified dissipation is no longer contradictory to the mass-budget closure in the bottom ocean.

The dichotomy of diffusivities can be also re-deciphered. According to the unsteady abyssal recipes, the diffusivity should match with the diapycnal velocity, $w_{dia}$, rather than the Eulerian velocity, $w_{Eul}$. Despite $w_{Eul}$ is constrained to a certain magnitude (e.g., 1×10$^{-7}$ m/s) subject to the mass-budget closure, the magnitude of $w_{dia}$ varies more flexibly. As an extreme example, $w_{dia}$ can even vanish if the motion is entirely adiabatic as the material surfaces move in sync with isopycnals at all times, leading to $w_{Eul} = w_{iso}$. In this scenario, diffusivity should be zero because adiabaticity implies no mixing. That is to say, there is no longer need to seek large diffusivities in the ocean with the goal of obtaining a global average of O(10$^{-4}$) m$^2$/s. The globally averaged diffusivity to balance the bottom-water formation rate does not have to be O($10^{-4}$) m$^2$/s. In another word, a locally vanished $\kappa$ does not prevent the water parcels from upwelling.

It is worth noting that the unsteady advection-diffusion equation is not new. However, in practical applications, the isopycnal velocity $w_{iso}$ is largely ignored. In such cases, the unsteady advection-diffusion equation is reduced to the steady one (compare Eqs. (1) and (3)).

Despite the conundrum seems to be reconciled, the exact physical processes driving the continuous ascending of bottom water remain obscure. The BBL upwelling theory (introduced in Section 1) offers one solution to this enigma. However, there are cases of abyssal upwelling that this theory is not applicable, such as for flat-bottom topography or non-bottom-intensified diffusivity. In the next section, these cases are investigated with a suite of idealized experiments. Based upon the results of the numerical experiments, a new conceptual model of abyssal upwelling is proposed, providing an alternative explanation to this fundamental question.

## 3. Numerical experiments on abyssal upwelling and a new conceptual model

### 3.1 Abyssal upwelling in the coarse-resolution simulations

For the BBL upwelling theory to operate effectively in the ocean models, it is necessary for the models to either resolve or parameterize the BBL layer. As noted earlier, this layer has a relatively narrow spatial extent, approximately 0.2° of longitude or latitude (McDougall and Ferrari 2017). However, many coarse-resolution ocean models have proven capable of successfully simulating abyssal upwelling. Table. 1 provides a selection of such model studies.

| Model ID | Region | Model resolution | Diffusivity $\kappa$ | Topography | Abyssal upwelling | Reference |
|---|---|---|---|---|---|---|
| 1 | Global | 3.75°×4.5° | BI | Realistic | 12 Sv | Toggweiler and Samuels (1998) |
| 2 | Pacific | 2.8°×2.8° | BI+U | Realistic | 8–10 Sv | Tsujino et al |

| | | | | | (2000) |
|---|---|---|---|---|---|
| 3 | Atlantic | 3.5°×3.5° | BI+U | Realistic | 6 Sv | Prange et al (2003) |
| 4 | Indo-Pacific | 3.8°×1.8° | BI | Realistic | 10.4 Sv | Simmons et al (2004) |
| 5 | Indo-Pacific | 1.87°×1.87° | BI | Realistic | 6 Sv | Saenko and Merryfield (2005) |
| 6 | Idealized basin | 2°×2° | U (1×10$^{-4}$ m$^2$/s) | Flat bottom | 7 Sv | Nikurashin and Vallis (2011) |
| 7 | Indo-Pacific | 1°× (1/3~1)° | BI | Realistic | 10 Sv | Melet et al (2013) |
| 8 | Idealized basin | 2°×2° | BD | Flat bottom | >16 Sv | Ferrari et al (2016) |

Table 1. A selection of coarse-resolution models that have simulated abyssal upwelling. Model resolution is expressed as longitude × latitude. "BI", "BD" and "U" in the column of "Diffusivity" represent bottom-intensified, bottom-decayed, and uniform vertical distribution of $\kappa$, respectively.

Models listed in Table 1 all generated a reasonable amount of abyssal upwelling, despite none of them had a horizontal resolution sufficient to resolve the BBL. Ferrari et al (2016) argued that models prescribing uniform diffusivity in the deep ocean is likely to produce a bottom-decreased density flux, which is inconsistent with the observations. Still, those simulations with uniform or even bottom-decayed diffusivity can produce an upwelling similar to others (e.g., Models 2, 3, 8 in Table 1). These results seemingly imply an upwelling mechanism that does not necessarily hinge on the vertical distribution pattern of diffusivity, the resolving of BBL, or even topographic characteristics. While these factors were considered indispensable in the BBL upwelling theory (e.g., de Lavergne et al 2016). Therefore, there might be a distinct mechanism that accounts for the upwelling in those cases.

### 3.2 Numerical experiments using box model

To investigate the cause of the abyssal upwelling in the coarse-resolution models, I conduct an idealized box simulation with a periodic or reentrant channel using MITgcm (Marshall et al 1997). Setup of this box-model provides a well-idealized representation of a closed basin connected to a laterally unbounded channel to the south. Thus, it has been favored in the mechanism study of the global MOC (e.g., Ito and Marshall 2008; Wolfe and Cessi 2010; Nikurashin and Vallis 2011; Wolfe and Cessi 2011; Munday et al 2013; Bell 2015; Mashayek et al 2015; Ferrari et al 2016). Considering that the model by Ferrari et al (2016) (hereafter FMM16) provides a particularly suitable example for the purpose here, I adopt its entire configuration for my control run, except for the mixing profiles (see supplementary Table S1 for a list of modeling parameters). The model domain of the control run is comprised of a reentrant channel at 60°–40°S and a rectangular flat-bottom basin from 40°S to 60°N (Fig. 3) with a resolution of 2°×2°. Following the setting of FMM16, the model is forced at the surface by restoring temperature to a prescribed profile, and wind stress that imposes above the open channel area only. The configuration file is amended based on one of the MITgcm tutorial example experiments, i.e., the "Southern Ocean Reentrant Channel Example" (accessed Dec 2023).

Even though FMM16 prescribed the profiles of density flux, it is a common practice to prescribe the vertical diapycnal diffusivity, $\kappa$, in most of the numerical experiments. The current experiment tests seven distinct vertical distributions of diffusivities. Using a similar box model, Nikurashin and Vallis (2011) found a strengthened abyssal overturning with greater magnitude of constant diffusivity. However, the impact of the vertical distribution of diffusivity has not been examined. Fig. 3 (right panel) shows the seven distributions of $\kappa$, among which Case 4 is the control run, with a vertically uniform diffusivity of $1\times10^{-4}$ m$^2$/s. On the other hand, Cases 3 and 6 represent a bottom-intensified diffusivity, while Cases 2 and 5 represent a bottom-attenuated diffusivity. All of the cases reach a quasi-steady state after running for 4000 years (The steady state is generally reached for most cases after 3000 years, a period consistent with Ito and Marshall (2008). While cases with weak diffusivity, such as Case 1, requires longer time to reach quasi-equilibrium). The diagnostics at year 4000 are selected for analysis of the final state.

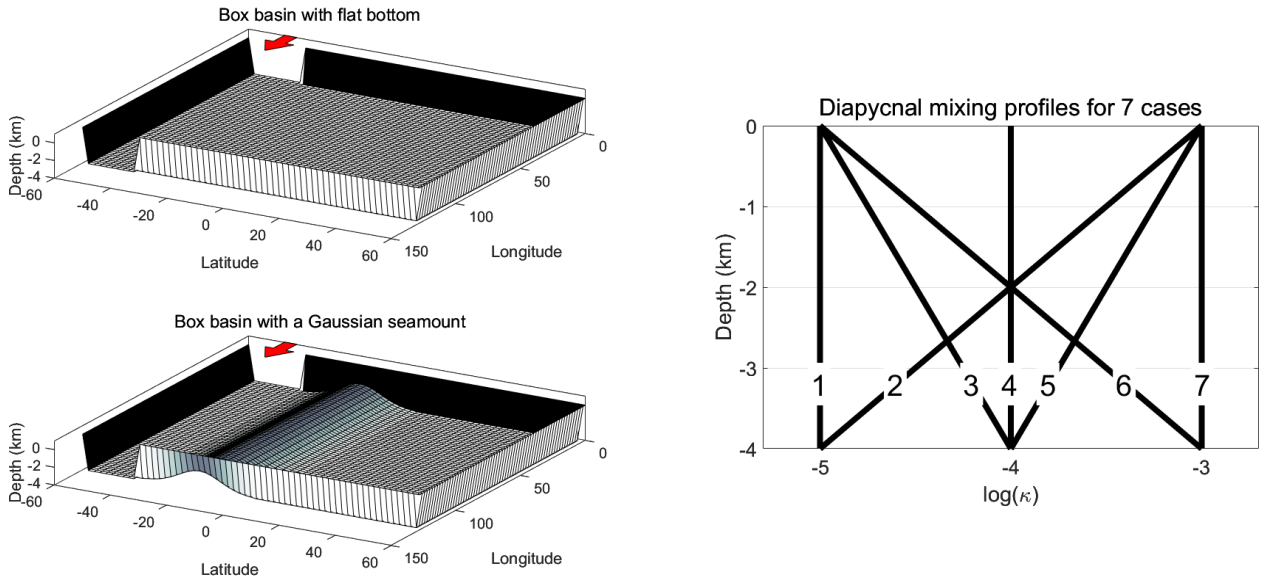

Fig. 3. Topography of the model domain (left) and profiles of diapycnal mixing for the 7 cases (right). The topography with a zonally-symmetric seamount is shown in the lower left. The red arrows indicate the direction of the zonally uniform wind that only imposes over the channel. In addition to the 7 cases of diffusivities, additional experiments based on the control run (Case 4) are conducted, including: Case 4-SOe3: increased diffusivity in the open channel to $10^{-3}$ m²/s; Case 4-SOe5: decreased diffusivity in the open channel to $10^{-5}$ m²/s; Case 4-mount: using seamount topography instead of the flat bottom; Case 4-nowind: removal of wind forcing. Refer to the text for discussions on all experimental results.

To evaluate the dependency of abyssal upwelling on the vertical distribution pattern of diffusivity ($\kappa$), the residual MOC streamfunctions are compared in Fig. 4 for three cases. The residual MOC is defined as the net effect of transport by the Eulerian-mean and the eddy-induced circulation (e.g., Marshall and Radko 2003; Nurser and Lee 2004; Ito and Marshall 2008; Munday et al 2013). The Eulerian and eddy-induced MOC streamfunctions are calculated with the Eulerian velocity and the parameterized bolus velocity, respectively (Gent and McWilliams 1990; Ito and Marshall 2008). In fact, the residual MOC does not differ much with the traditional Eulerian MOC in the interior of the basin, which is the domain of interest of this study (Ito and Marshall 2008; Fig. S1). It is because the eddy-induced MOC is relatively small due to the rather flat isopycnals therein.

The first conclusion that can be drawn from Fig. 4 is, consistent with Table 1, abyssal upwelling (or equivalently, abyssal overturning circulation) can be well produced in the interior of the basin, even with a flat-bottom topography. Secondly and more importantly, among the three cases compared, the upwelling does not increase with a bottom-intensified diffusivity (Case 2). Rather, it is largest with a bottom-decayed profile (Case 5). This comparison seems to suggest that the abyssal upwelling may not depend on the vertical distribution of the diffusivity, but rather on the mixing intensity (magnitude of $\kappa$) itself in the coarse-resolution model. The residual MOC streamfunctions for the other four cases are shown in Fig. S2 for reference. To facilitate a quantitative comparison on the upwelling intensity, an abyssal MOC index is defined with the opposite MOC value at a depth of 3000 m and a latitude of 10°S. This index represents how much water to the north of 10°S upwells/overturns across the depth of 3000 m per unit time. The values of this index are summarized in Table 2 for all cases. As noted earlier, the difference between the residual and the Eulerian MOC streamfunctions is generally small. Results of Table 2 further affirm that the abyssal overturning strength mostly depend on the magnitude, rather than vertical distribution of $\kappa$ in the deep ocean. Two additional experiments (Case 4-SOe5 and Case 4-SOe3) are conducted with a different diffusivity prescribed for the open channel area: $1\times10^{-5}$ m$^2$/s and $1\times10^{-3}$ m$^2$/s, respectively. Though not altering the above conclusion significantly, their results indicate that higher diffusivity in the "Southern Ocean" tends to cause a reduction in upwelling intensity within the interior basin (Table 2). This phenomenon may be attributed to the fact that the densest water, formed at high latitudes, is less able to intrude into the interior basin due to being weakened by stronger mixing. This result supports the conceptual model to be proposed at a later stage.

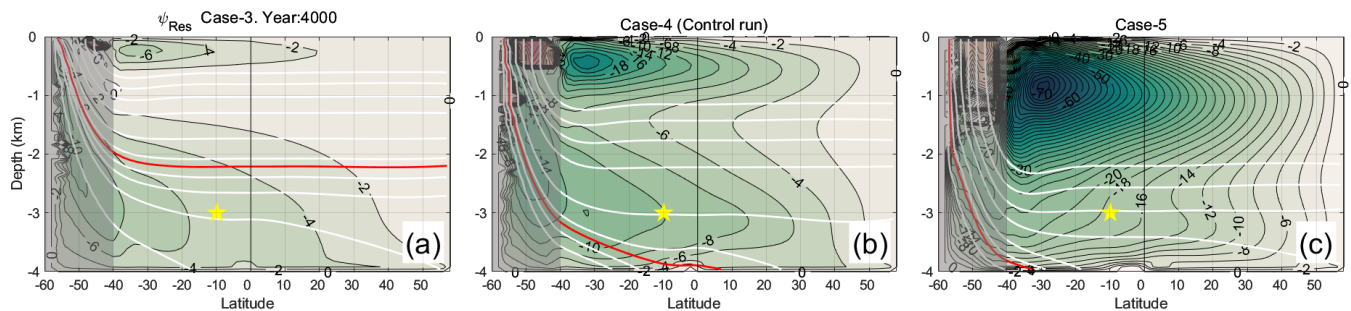

Fig. 4. The residual MOC streamfunctions (unit: Sv) with the prescribed diffusivity in (a) case 3, (b) case 4, and (c) case 5 in Fig. 3. Negative streamfunction indicates anti-

clockwise overturning cell. Temperature contours are depicted by the white lines, with the red line highlighting the 3°C contour. The yellow star marks the position used to define the abyssal upwelling index. The open channel domain is shaded.

Concerning the vertical velocity, FMM16 showed the diapycnal velocities, which was computed with the prescribed turbulent density flux. Here, we present a more straightforward indicator of water parcel vertical movement: the Eulerian vertical velocity, which is a direct output from MITgcm. Fig. 5 shows this component in 2D sections for the control run. Except the western boundary, upwelling prevails in the entire abyss of the interior basin. These results are consistent with the overturning streamfunctions in Fig. 4 and Fig. S2. An intriguing phenomenon (though less relevant to the topic) is that the bottom water intrusion, which can be indicated by the green-dashed isotherm, tends to intensify towards the western boundary. Similar patterns of upwelling are found in other cases as well. The reason is unclear.

| Case ID | 1 | 2 | 3 | 4 (control) | 5 | 6 | 7 |
| --- | --- | --- | --- | --- | --- | --- | --- |
| Residual MOC (10°S) | 0.05 | 5.4 | 8.3 | 10.1 (10.3; 8.4) | 13.8 | 17.4 | 49.8 |
| Eulerian MOC (10°S) | 0.1 | 4.9 | 8.3 | 10.0 (10.3; 8.4) | 12.8 | 17.3 | 49.9 |
| Residual MOC (10°N) | 0.0 | 6.1 | 4.5 | 8.5 (8.7; 7.4) | 14.2 | 11.2 | 44.7 |
| Eulerian MOC (10°N) | 0.1 | 6.0 | 3.5 | 8.4 (8.6; 7.5) | 14.4 | 7.8 | 43.7 |

Table 2. Abyssal upwelling indexes for the seven MITgcm experiments (unit: Sv). See text for the definition of the index. The numerical cases are arranged by the magnitude of deep diffusivity below 2000 m, from smaller to larger (see Fig. 3). Cases 2 and 3, as well as 5 and 6, are grouped together because the magnitudes of their deep diffusivity cannot be definitely separated. The values in the parentheses of the Case-4 column denote the results from two additional numerical experiments prescribing different diffusivities in the "Southern Ocean" (Case 4-SOe5 and Case 4-SOe3). The upwelling indexes evaluated at 10°N are also shown in the last two rows.

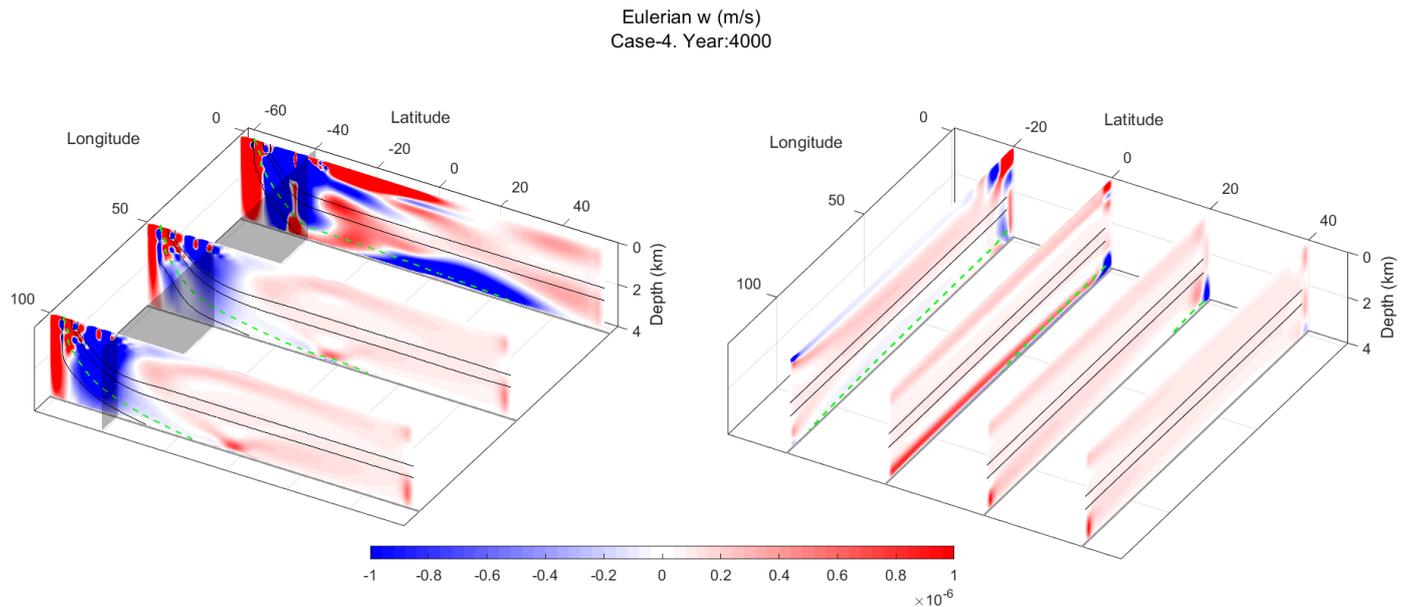

Fig. 5. The meridional (left) and zonal (right) sections of the Eulerian vertical velocity (color shading) for the control run of MITgcm (Case 4). Also superimposed are isotherms (black curves), with the 3°C one highlighted in green dashed. The transparent vertical plane in the left panel marks the northern boundary of the open channel.

Stronger mixing corresponds to stronger upwelling. This conclusion was identified in some previous modeling studies as well. But few studies have attempted to isolate the influence of local mixing and bottom water intrusion from the Southern Ocean. Here, another experiment is conducted using seamount topography to obstruct the intrusion of bottom water into the basin interior (Fig. 3). This case is referred to as "Case 4-mount" as it shares the same configuration with the control run, except for the bottom topography. Fig. 6 compares the development of the abyssal overturning circulation for the two cases. By Year 1000, the abyssal MOC has already been well established in Case 4, while the abyssal ocean below 3000 m in Case 4-mount remains quiescent even though the mixing in the northern hemisphere of the two cases is exactly the same. The water below 3000 m in Case 4-mount is only upwelled when the densest water south of the seamount continues to accumulate and finally spills out over the underwater dam. A strong downslope current on the northern face of the seamount is more pronounced in the Eulerian vertical velocity by Year 4000 than in Year 1000, particularly in the western half of the basin (Fig. S3, S4).

The evolution of the abyssal MOC index for Case 4 and Case 4-mount until Year

4000 is illustrated in Fig. S5. While the two cases have reached similar final upwelling intensity, the establishment of the abyssal MOC in the seamount case is significantly delayed by approximately 1000 years compared to the control run. An additional case, also depicted in Fig. S5, examines the impact of wind forcing (Case 4-nowind). Interestingly, it is observed that the absence of wind has minimal influence on the intensity of the abyssal MOC. Does this suggest that the abyssal MOC is primarily driven by buoyancy forcing, while the Antarctic Circumpolar Current and the westerlies are dispensable in this process? This question is beyond the scope of this paper but warrants detailed exploration in future research.

These results suggest that abyssal upwelling cannot be achieved through mixing alone. The intrusion of the densest bottom water is another indispensable factor in driving the upwelling process in the abyss. The results from Case 4-SOe5 and Case 4-SOe3 in Section 3.2 support this role of bottom water by demonstrating that a weakened AABW leads to a weakened abyssal overturning in the basin interior. The upwelling of abyssal water is the result of the combined effects of mixing and bottom-water intrusion.

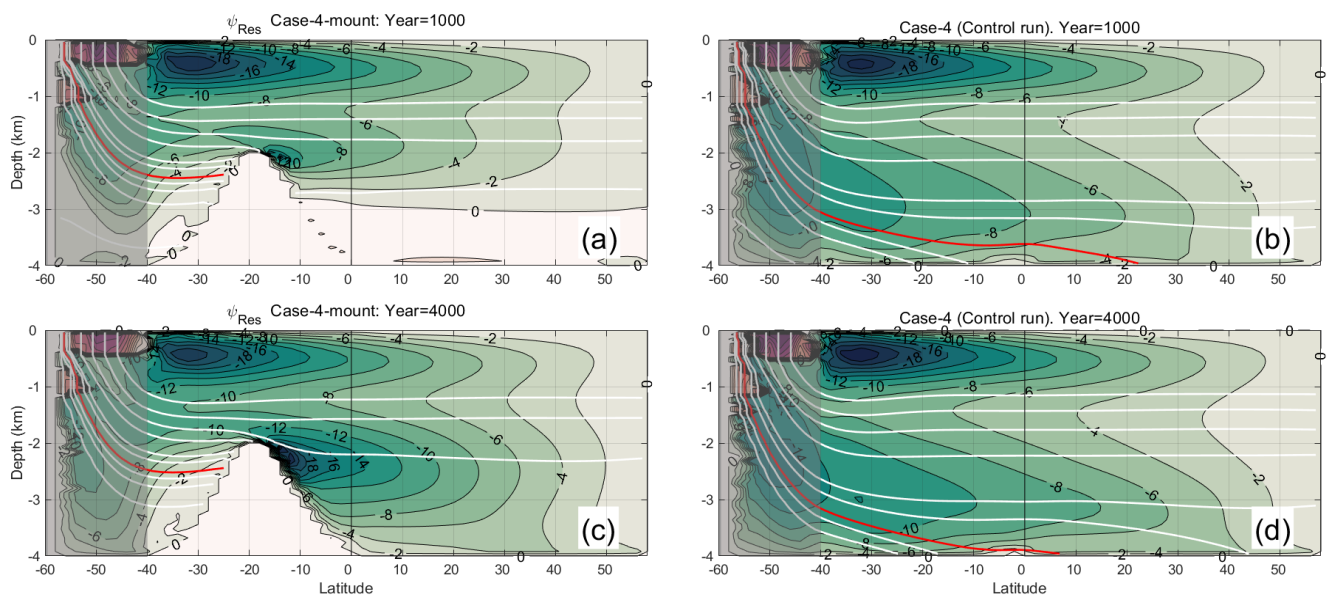

Fig. 6. The same as Fig. 4, but for Case 4-mount (a, c) and Case 4 (b, d), and for Year 1000 (a, b) and Year 4000 (c, d).

### 3.3. An erosion-intrusion model of bottom-water upwelling

A conceptual model for the abyssal upwelling can be envisioned based upon the results of the coarse-resolution numerical experiments with a flat bottom. This model, as illustrated in Fig. 7, comprises four processes: erosion, transformation, intrusion and expansion. Detailed explanations for each stage are provided in the caption of Fig. 7. The 2 °C-isotherm demarcates the border of AABW in the Indo-Pacific Ocean and the 3 °C-isotherm signifies an adjacent isotherm positioned above the 2 °C-isotherm. The specific temperature values are not essential, but are used for narrative convenience. The dynamical reason why intrusion is induced by the erosion of the AABW front is straightforward: The pressure at the original position of AABW front decreases as the density is reduced by the continuous downward heat flux. This leads to restoration of the pressure balance, assuming the pressure head at the upstream of AABW is generally maintained throughout the process. As the four processes continue (with the downward heat flux continuously eroding the bottom water), parcel A is continuously pushed up by the persistent intrusion of AABW. Density-driven flow, or the gravity current (e.g., the downslope current in Case 4-mount), can be accurately simulated by numerical ocean models or climate models (e.g., Legg et al 2009; Laanaia et al 2010; Reckinger et al 2015). Considering the extensive scale of bottom intrusion flow, even models with coarse resolutions can effectively resolve it. Hence, the conceptual model depicted in Fig. 7 offers an explanation for the effectiveness of coarse-resolution models in generating abyssal upwelling.

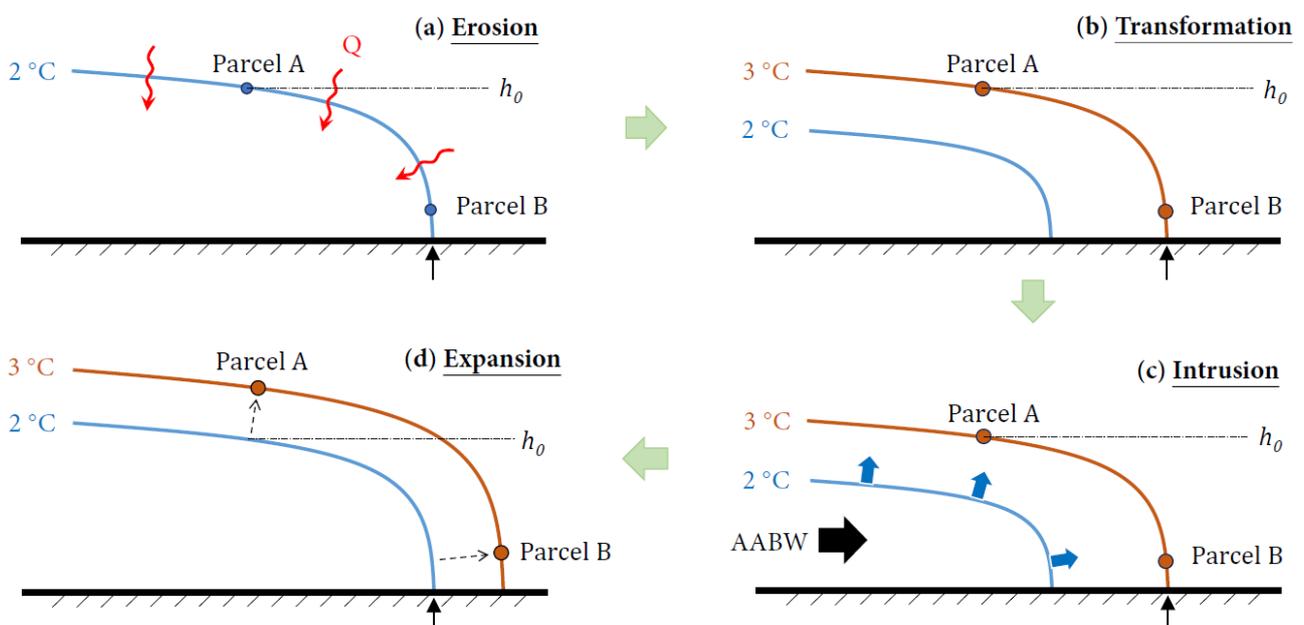

Fig. 7. Schematic diagram illustrating the erosion-intrusion model of abyssal upwelling.

The blue and brown curves indicate isotherms of 2°C and 3°C, respectively. Initially, the bottom water, characterized by temperature below 2°C, stays in a balanced location. Its front intersects the seafloor (at a point marked by a small black arrow) in all four processes. Two water parcels A and B reside at different locations on the 2 °C-isotherm. The upwelling of parcel A from its original depth (indicated by $h_0$) occurs through four processes: (a) Erosion: the downward heat flux due to mixing warms the coldest bottom water as the seafloor is insulated; (b) Transformation: as the bottom water is warmed, the bottom-water front contracts while the temperatures in parcels A and B being increased to, e.g., 3°C; (c) Intrusion: since the initial balance for the bottom-water front is disrupted, there is a tendency for the bottom water to recover its original extent, being pushed by the newer/younger bottom water from the source region; (d) Expansion: intrusion of the newer/younger bottom water pushes all the water parcels further away. Parcel A is pushed to a higher position (or a shallower depth) and parcel B is pushed further north. Parcel A and B are not necessarily situated at the same isopycnal throughout the process. Note this schematic illustrates a steady abyss for simplicity of representation, but it is equally applicable to scenarios involving cooling or warming abysses.

Warming of the bottommost water is independent on the vertical divergence of turbulent heat flux because it is the coldest and the seafloor is thermal-insulated. In cases where the seafloor is not insulated due to the presence of geothermal vents, the bottom water is warmed even more rapidly. The four processes are depicted sequentially in Fig. 3 for the sake of simplicity and clarity in illustration. In reality, the erosion and intrusion processes occur simultaneously, suggesting that the AABW front is in a dynamic equilibrium.

According to the erosion-intrusion model, the main factor determining the vigor of bottom-water upwelling is not the vertical structure of mixing but rather the intensity of the mixing itself. This is because stronger mixing accelerates the erosion process, which in turn facilitates a more rapid intrusion of AABW. In their global runs with uniform diffusivities (Model 1 in Table 1), Toggweiler and Samuels (1998) found that the upwelling transport increased nearly linearly with the elevated diffusivities. However, their results were presented only for a depth of 1350 m, rather than at an abyssal level. Using a flat-bottom model (Model 6 in Table 1), Nikurashin and Vallis (2011) similarly demonstrated that the deep overturning cell strengthened with increasing diffusivity. In a study using a set of coarse-resolution circulation models that incorporated radiocarbon tracers, Koeve et al (2015) observed a decrease in radiocarbon age (indicating water becoming younger) in the deep North Pacific as the background mixing coefficients increased, suggesting the deep ocean being refreshed by a faster overturning circulation. In a recent study, Kawasaki et al (2022) tracked the upwelling of Lagrangian particles in the deep North Pacific using a 1°-resolution model. In their "case 3D", where horizontally non-uniform diffusivity is prescribed using a high-resolution tide model, the ascending hotspots of the abyssal particles are concentrated around ridges, seamounts, and boundaries. Conversely, in their "case 1D", where diffusivity is horizontally uniform and also bottom-intensified, these hotspots are distributed homogeneously horizontally. While these two cases cannot be explained by the BBL theory due to the insufficient model resolution, they can be explained by the erosion-intrusion model: In case 3D, the strong mixing that occurs within topographic features leads to rapid erosion of the bottom water, resulting in increased upwelling in those regions. In contrast, case 1D features uniform mixing, leading to a uniformly distributed erosion of bottom water and consequently, a horizontally uniform upwelling throughout the basin. Eddy-resolving models have also identified a significant enhancement of upwelling at a depth above the topography (thus ruling out BBL upwelling) at energetic eddy hotspots (Tamsitt et al 2017).

It is worth noting that the water parcel away from the seafloor can also rise, even in the absence of local mixing. This ascent is facilitated by the continuous push exerted by the intruded bottom water from below. This scenario highlights that Munk's abyssal recipes are not directly linked to the upwelling of water parcels, despite this being the theory's original objective. Strictly speaking, Munk's theory describes the decoupling rate between the material surface and property surface (isopycnal or isotherm). In simpler terms, it explains how fast properties (e.g., temperature) on the material surface are transformed by diabatic processes (e.g., heat flux due to mixing). The actual rising of water parcels is not contingent on the formula.

The rising pattern of bottom water in the erosion-intrusion model resembles the growth of "*tree rings*". Let's assume the group of water parcels that are located on the border surface of AABW (e.g., the 2 °C-isotherm in Fig. 7) have approximately the same age (water age is conventionally defined as the elapsed time since the last contact with the atmosphere). As time passes, one after another group of water parcels with the same ages is sequentially pushed up and moves farther away from the seafloor. These water parcels undergo aging as they ascend. Therefore, tracers with a temporal decay property (e.g., dissolved oxygen and radioactive substances) may expose the age distributions of water parcels, similar to the pattern observed in tree rings. In the next section, the ascent rate of abyssal water will be estimated based on this concept of "*abyssal tree rings*".

## 4. Estimating upwelling speed in the deep North Pacific using biogeochemical data

In this section, data from the Global Ocean Data Analysis Project (GLODAP) is employed to estimate the Eulerian rising velocity. GLODAP is a comprehensive effort to synthesize biogeochemical data collected from ocean surface to bottom by analyzing water samples for chemical composition (Olsen et al 2016). The version used in this study is GLODAPv2.2022, which includes 13 core variables derived from over 1.4 million water samples collected in 1,085 research cruises up to 2021 (Lauvset et al 2022). A subset of this dataset is used, specifically focusing on the deep Pacific Ocean (>3000 m) north of 10°S, where AABW enters the basin through the Samoan Passage (Hogg et al 1982; Roemmich et al 1996; Voet et al 2016). In the following, two biogeochemical tracers, dissolved oxygen (DO) and radiocarbon abundance ($\Delta^{14}C$), are chosen for analysis.

### 4.1 Abyssal recipes for dissolved oxygen

The primary consumption process of the oxygen in the deep ocean is remineralization of organic matter, which can be viewed as the reverse of the photosynthesis reaction. If remineralization were to dominate distribution of oxygen and phosphate (P), one would expect a linear relationship between the two tracers, with a ratio of DO:P=−170:1 (Sarmiento and Gruber 2006). According to the GLODAP data, a linear regression between DO and phosphate finds a slope of −164 with a coefficient of determination of 0.89 in the deep Pacific Ocean (Fig. 8a). This strongly suggests that remineralization is the predominant process of DO consumption in these waters. Hence, the consumption of DO in the deep Pacific Ocean can be described by the oxygen utilization rate (OUR) linked to the remineralization of organic matters. This rate is defined as the material derivative of oxygen with respect to time and is typically treated as a constant within the same basin (Chapter 5 of Sarmiento and Gruber 2006).

Similar to Eq. (3), the unsteady advection-diffusion-decay balance of DO can be expressed as follows,

$$\left(w_{Eul} - w_{iso}^{DO}\right)\frac{\partial [DO]}{\partial z} = \frac{\partial}{\partial z}\left(\kappa \frac{\partial [DO]}{\partial z}\right) - \text{OUR}, \tag{6}$$

where $[DO]$ denotes the concentration of DO and $w_{iso}^{DO}$ the vertical displacement velocity of DO-isosurface. The concentration of DO increases towards the ocean bottom, with contours approximately parallel to the front of the bottom water, as observed in hydrographic sections (e.g., Talley 2013). This distribution is qualitatively consistent with the "tree ring" concept inferred from the erosion-intrusion model. When examining the vertical distribution of DO by aligning all the profiles to their bottommost values, it becomes evident that the distributions are primarily linear (Fig. 4b). If we consider the rising at a timescale within a limited time (e.g. a century), the linear distribution becomes even more pronounced within a vertical extent of O(10) m, given the rising velocity on the order of O(1) m/year. Furthermore, if the diffusivity is considered uniform within the extent of O(10) m, the diffusion term in the RHS of Eq. (6) is close to zero. Therefore, it is neglected in the balance equation in the following estimation. The impact of a bottom-intensified diffusivity will be discussed in the next subsection.

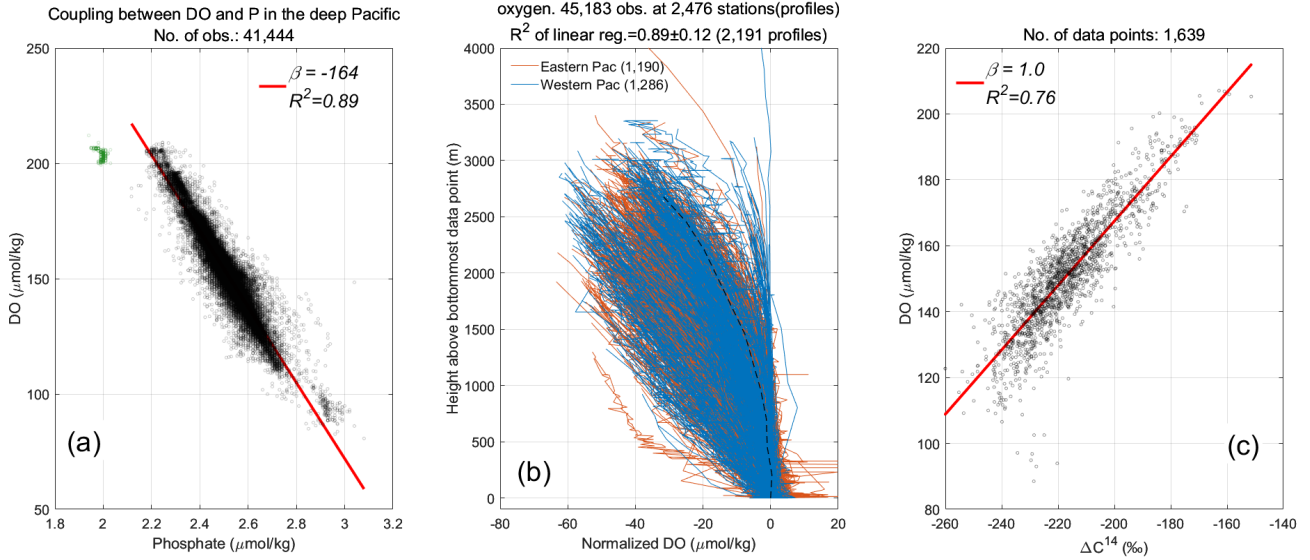

Fig. 8. The properties of dissolved oxygen (DO) in the deep Pacific Ocean (>3000 m) north of 10°S. (a) Linear regression between DO and phosphate with a total number of 41,444 data points. The linear slope ($\beta$) and coefficient of determination ($R^2$) are -164 and 0.89, respectively. The cluster of outliers (green) with the lowest phosphate concentration is from the Japan Sea and has been excluded in the linear regression analysis. (b) Normalized DO by subtracting from each data point the value of the bottommost data point of each station/profile. There are a total of 2,476 stations/profiles involved, with 1,190 from eastern Pacific and 1,286 from western Pacific, divided by the 180°-meridian. Data from multiple (repeat) cruises at the same location are merged into one profile. Linear regression analysis on each profile yields coefficients of determination in the range of 0.89±0.12 for the 2,192 profiles (profiles with fewer than four data points are excluded). The black dashed curve is an example profile with bottom-decreased vertical gradient. The difference in number of observations between (a) and (b) is due to no phosphate data in 3,427 water samples and exclusion of the anomalous low phosphate data in the Japan Sea (312 water samples). (c) Linear relationship between DO and radiocarbon for all the 1,639 available measurements. Data: GLODAP v2.2022.

Next, let's examine the LHS of Eq. (6). Depletion of DO at abyssal Pacific was identified by decadal time-scale observations (Whitney et al 2007; Smith Jr et al 2022). Since DO increases towards the bottom in the Pacific, declining trend in DO implies a descending DO-isosurface, i.e., a negative $w_{iso}^{DO}$. As a result, the balance equation can be rewritten as follows,

$$\left(w_{Eul} + |w_{iso}^{DO}|\right)\frac{\partial[DO]}{\partial z} \approx -\text{OUR}, \qquad (7)$$

Note the two vertical velocity components in the advection term are both positive.

**4.2 Estimating rising speed with water age from radiocarbon tracer $\Delta^{14}$C**

The radiocarbon $^{14}$C is a valuable tracer for understanding the movement and mixing of carbon in the ocean. $\Delta^{14}$C is a measure of the relative abundance of the radioactive isotope $^{14}$C in a sample of carbon compared to a reference standard, typically expressed in per mil (‰). It offers valuable information about the age and origin of carbon within the ocean and can be employed to estimate the age of water masses. Lower $\Delta^{14}$C values typically indicate older water, implying that it has not been in contact with the atmosphere (where $^{14}$C is produced) for an extended period (Sarmiento and Gruber 2006).

The $\Delta^{14}$C distribution in the hydrographic sections in the deep Pacific exhibits similar pattern as DO, i.e., with the contours approximately parallel to the front of bottom water (e.g., Key 2001; Matsumoto and Key 2004; Jahn et al 2014; Koeve et al 2015). The pattern of $\Delta^{14}$C bears more resemblance to that of a tree ring because it can be converted to the age of water (in years) by the formula (Stuiver and Polach 1977; Matsumoto 2007; Koeve et al 2015),

$$^{14}\text{C age} = -8033 \cdot \ln\left(1 + \frac{\Delta^{14}\text{C}}{1000}\right). \tag{8}$$

This formula suggests an approximate difference of 10 years in age per unit change of $\Delta^{14}$C (in mil) within the range observed in the deep Pacific Ocean (Fig. 8c).

Relating the DO and $\Delta^{14}$C yields a linear slope of 1.0 μmol/kg/‰ with $R^2$=0.76 (Fig. 8c). The OUR can be thus calculated as 1.0 (μmol/kg) /10 (years) = 0.1 μmol/kg/year. This value compares with the former estimates of 0.13 μmol/kg/year (below 2000 m, Chen 1990) and 0.14 μmol/kg/year (below 1500 m, Sarmiento and Gruber 2006) in the Pacific. For readers' reference, if the depth limit in my filtering is changed from below 3000 m to below 1500 m, an elevated OUR of 0.17 μmol/kg/year is obtained, but with a reduced $R^2$ of 0.5.

Given the significant correlation between DO and $\Delta^{14}$C, as well as the similar bottom-elevated distribution of $\Delta^{14}$C, it is reasonable to assume that the $\Delta^{14}$C - isosurface in the deep Pacific descends as the DO-isosurface does. Hence, a similar balance equation for $\Delta^{14}$C can be obtained,

$$\left(w_{Eul} + \left|w_{iso}^{C14}\right|\right)\frac{\partial[\Delta^{14}C]}{\partial z} \approx -\text{RDR}, \tag{9}$$

where $w_{iso}^{C14}$ denotes the vertical velocity of the $\Delta^{14}$C-isosurface and RDR the "radiocarbon decay rate", which has a value of 0.1 ‰/year as previously estimated.

Now, with the vertical profiles of DO and $\Delta^{14}$C, the cross-isosurface vertical velocity components in Eqs. (8) and (9), i.e., $w_{Eul} + \left|w_{iso}^{DO}\right|$ and $w_{Eul} + \left|w_{iso}^{C14}\right|$, can be estimated, respectively, for each station in the GLODAP dataset. Only profiles with a linear coefficient of determination ($R^2$) greater than 0.7 are included in the calculation. As a result, 2,157 stations for DO and 179 stations for $\Delta^{14}$C in the deep Pacific Ocean north of 10°S are qualified for the estimation.

The results are shown in Fig. 9. Due to the addition of a non-negative term, all the upwelling velocities depicted in Fig. 9 represent an upper limit of the Eulerian upwelling velocity, $w_{Eul}$. There are both agreements and disagreements between the two estimates from DO and $\Delta^{14}$C in the details of spatial distributions. However, there is a broad consensus that the vertical velocities tend to be higher in the western and northern areas. This qualitatively aligns with the recent numerical study using Lagrangian particles, which suggests that "upwelling of deep water is confined in the western North Pacific Ocean owing to the strong vertical mixing" (Kawasaki et al 2022). Another study also noticed a notably high positive Eulerian vertical velocity in the northern end of the Pacific across the depth of 3000 m using a reanalysis product (Liang et al 2017).

The estimated vertical velocity mostly remains below 10 m/year, with the peak of the distribution occurring within the range of approximately 4−9 m/year (Fig. 9c). The Eulerian vertical velocity below 2500 m in the Northern Hemisphere of the MITgcm experiment is also shown (Fig. 9d). It exhibits a peak at a lower velocity, around 3 m/year. These two estimates are of the same magnitude and align with the earlier assertion that the 4−9 m/year estimate represents an upper limit for the Eulerian vertical velocity, $w_{Eul}$.

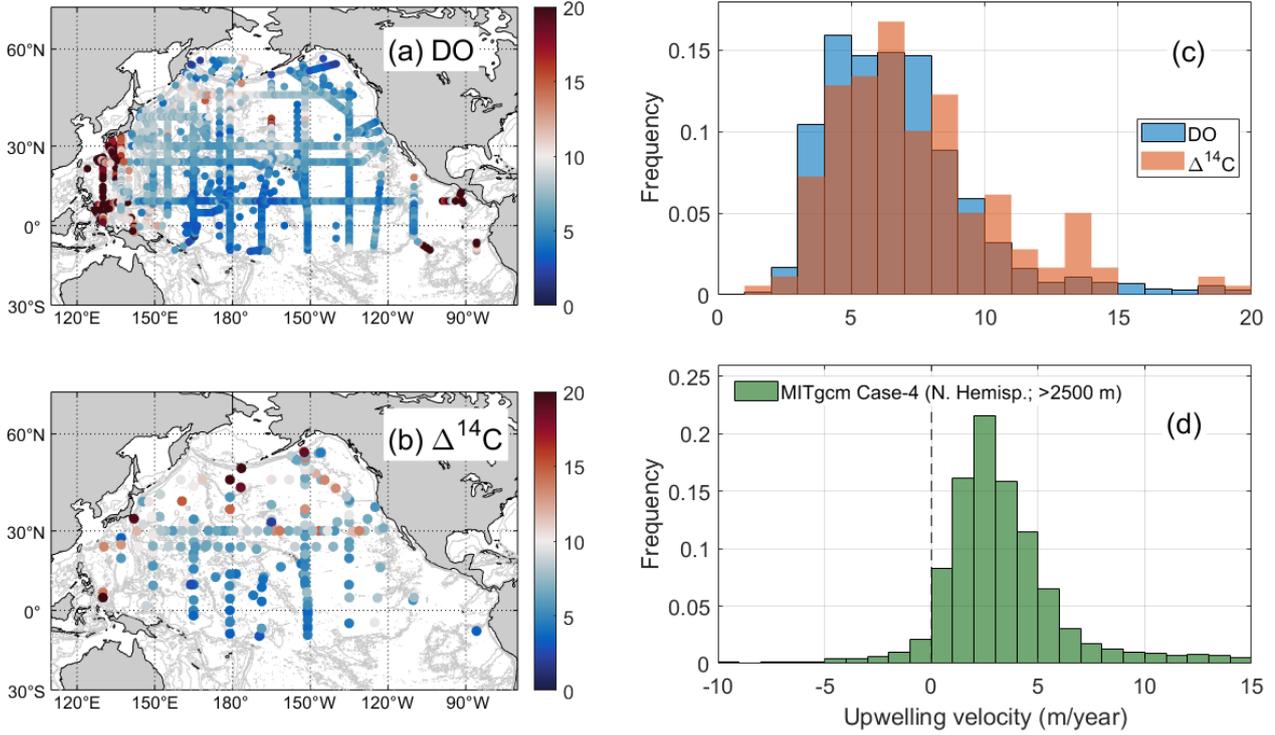

Fig. 9. The mean upwelling velocity (relative to the isosurface) below 3000 m in the North Pacific estimated using data of DO (a) and $\Delta^{14}C$ (b). Colors indicate the upwelling magnitude in m/year: Warm (cold) color denotes an upwelling velocity above (below) 10 m/year. Depth contours of 6000 m, 4000 m, and 2000 m are superimposed in grey curves to show topographic features. (c) Distribution of the upwelling velocities in (a) and (b). Profiles with linear coefficient of determination lower than 0.7 are not included. An OUR value of 0.1 μmol/kg/year is used in the calculation. Data: GLODAP v2.2022. (d) Same as (c), but for the MITgcm run of Case 4 (control run). Data involved in the statistics are the Eulerian vertical velocity below 2500 m in the Northern Hemisphere.

As mentioned in Section 2, the vertical advection-diffusion balance can be converted to a vector form. Indeed, according to the erosion-intrusion model, the expansion does not only occur in the vertical direction, but in the horizontal direction (e.g., parcel B in Fig. 3). The horizontal expansion velocity can be estimated with a smoothed distribution of $\Delta^{14}C$ in the meridional section (e.g., Key 2001; Matsumoto and Key 2004; Jahn et al 2014). Fitting these distributions in the deep Pacific in both vertical and horizontal directions yields a vertical velocity of 4.1−9.8 m/year and a northward velocity of 11−20 km/year. The scale ratio between the two expansion speeds is on the order of O(1000), close to the aspect ratio of the ocean. These proportional speeds serve as additional evidence of the existence of the expansion process. Moreover, it is found that the vertical velocity obtained with these smoothed fields increase towards the north, which is consistent with the spatial characteristics observed in Fig. 9. According to the section distributions depicted in Key (2001), the vertical gradient of $\Delta^{14}C$ in the deep Indian Ocean appears to be smaller than that in the deep Pacific Ocean, suggesting a possibility that bottom water in the Indian Ocean rises faster than that in the Pacific Ocean assuming the displacement trends of isosurfaces in the two basins do not differ much.

In addition to the distributions of DO and $\Delta^{14}C$, other variables exhibit a similar "tree ring" distribution in the deep Pacific due to their significant correlation with DO, which is subject to remineralization process. Apart from phosphate (Fig. 8a), nitrate and silicate also display robust linear relationships with DO, with linear coefficients of determination of 0.92 and 0.72, respectively, based on the GLODAP data (not shown). Thus, the distribution pattern of phosphate, nitrate, and silicate in the deep ocean are likely formed by the erosion-intrusion-related upwelling process as well.

**4.3 Uncertainty of the estimation**

It is worth noting that the estimate in Fig. 9 is an average from the water column below 3000 m. Since the vertical balance equation holds pointwise (Drake et al 2020; Rogers et al 2023; Wunsch 2023), the rising speed is not vertically uniform. Take a profile with bottom-attenuated gradient of DO for example (black dashed curve in Fig. 8b), the vertical velocity would decrease away from the bottom according to Eq. (7). Even though the vertical profile of rising speed is not determined here, it is highly likely that water parcels slow down as they rise to the intermediate depths. The Eulerian MOC streamfunctions indicate that bottom upwelling reaches a minimum at the depth of ~3000 m and water parcels turn to move horizontally southward, exiting the basin (Fig. 2a).

The value of OUR adopted in the estimation also introduces uncertainty. If we use an OUR value of 0.14 instead of 0.1 μmol/kg/year, the peak of DO-based estimate in Fig. 9c will shift slightly to about 5−11 m/year (not shown).

Recall that the diffusion term has been neglected in the estimation by assuming uniform diffusivity. However, if the diffusivity is notably bottom-intensified within the vertical extent of interest, it would result in a positive diffusion term for both DO and $\Delta^{14}C$ as per Eq. (6) ($\Delta^{14}C$ is not a concentration but a relative abundance, and thus its diffusion may not follow the conventional form as in Eq. (6) rigorously). This leads to a reduction in the upper limit estimate of the Eulerian rising speed.

## 5. Summary and discussions

This study revisits Walter Munk's seminal work on the abyssal upwelling in 1966 and investigates the factors that have posed challenges to Munk's theory in light of direct observations spanning decades since then. The steady-state assumption, as employed by Munk and his collaborators in the vertical balance equation, has been identified responsible for the paradoxes. Despite the adoption of the unsteady form in some later researches, omitting the displacement velocity of isopycnals in the interior ocean again reduces the unsteady form to the steady one.

Evidence suggests that the abyssal Indo-Pacific Ocean is experiencing a cooling phase, attributing to the cooling of the Antarctic occurred during the last Little Ice Age, which peaked centuries ago. This abyssal cooling, in line with the bottom-intensified mixing, implies upward displaced isopycnals. As a result, the diapycnal (rather than Eulerian) vertical velocity becomes negative (the so-called "interior downwelling"), which agrees with the observed bottom-intensified mixing (Eq. (5)). The conundrum in Munk's theory is thereby clarified by acknowledging that the theory does not address the vertical movement of water parcels. Instead, it describes the decoupling rate of the isopycnals and the material surfaces owing to diabatic processes.

A rather relieving inference is that it is no longer an imperative to seek a globally averaged diapycnal diffusivity of the order $O(10^{-4})$ m$^2$/s in order to close the mass budget associated with the bottom water formation rate. They are not supposed to correspond with each other. Mixing is not necessarily the local driving force of rising. A locally vanished diffusivity does not prevent the water parcels from upwelling.

Numerical experiments find that the abyssal upwelling as simulated in the coarse-resolution models unlikely depends on the bottom topography or vertical distribution of diffusivity. Instead, intrusion of bottom water and bottom mixing are both indispensable in driving the upward motions in the abyss. Based on this recognition, I propose an "erosion-intrusion" model with straightforward dynamics that can be adequately reproduced in the coarse-resolution models. This model operates on the bottom-water front with a straightforward conceptual framework. As the turbulent heat flux erodes the bottom-water front, the disruption in pressure balance pushes the upstream portion of the bottom water downstream to the original position of balance. This process continues like a "tug of war" between the mixing force and influx of younger bottom water from the Southern Ocean, ultimately resulting in a continuous process uplifting the older bottom water. Though the erosion-intrusion balance is proposed to explain abyssal overturning circulations, it may also apply to upper overturning cell, such as the Atlantic Meridional Overturning Circulation.

Using the observational data of the dissolved oxygen (DO) and radiocarbon abundance ($\Delta^{14}C$), I estimate the mean rising velocity below 3000 m in the North Pacific Ocean. As the vertical velocity of isosurfaces of both tracers cannot be determined with sufficient accuracy, this estimate, with a distribution centered in the range of 4−9 m/year, serves as an upper limit of the Eulerian rising velocity only. The distribution exhibits remarkable spatial non-uniformity, with higher rising velocities observed in the western and northern Pacific regions. In line with the erosion-intrusion model, these tracers (along with several others related to DO, such as nitrate, phosphate, and silicate) have been observed to exhibit characteristic distributions in the abyssal ocean with contours roughly parallel to the front surface of the bottom water. This pattern can be likened to "abyssal tree rings" due to their representation of water ages.

The key points of this paper can be duly summarized by addressing those thought-provoking and forward-thinking questions raised in the Abyssal Recipes II by Munk and Wunsch (1998) that are pertinent to this study:

1) *Their question 1: "Can enhanced boundary mixing, including that in canyons and along interior boundaries such as ridges and seamounts, balance the global formation of deep water in maintaining a steady abyssal stratification?"*

Firstly, the abyssal stratification is mostly unsteady. Assuming steady-state in the vertical advection-diffusion balance can lead to the conundrum of interior downwelling. Secondly, it is no longer necessary for the diapycnal diffusivity to balance the formation rate of deep water.

2) *Their question 5*: *"What is the connection between the strength of the convectively driven elements of the ocean circulation and the rate at which it mixes?"* (A question of focal point summarizing the new insights gained from this study)

   A possible way of connection is the one illustrated in the erosion-intrusion model (Fig. 7). These two components, symbolizing the actions of pull and push, function like two paddles in the bucket of MOC. The abyssal MOC, which represents the rate at which bottom water is refreshed, could be slowed down either by a weakened erosion rate due to reduced mixing or by a sluggish bottom-water formation. The question is, within the currently balanced overturning system in our global ocean, which component acts as the limiting factor? I would speculate that it is likely the mixing part (the pull force), as indicated by a model study that has identified a consistent trend of deep water becoming younger as background diffusivities increase (Koeve et al 2015). Another piece of evidence is the comparison between the numerical experiments, Case 4 and Case 7: With the bottom water formation process unchanged, the abyssal upwelling continues to intensify as the constant diffusivity increases by one order of magnitude. Considering that the averaged bottom diffusivity in the real world is very likely below $1\times10^{-3}$ m$^2$/s (Waterhouse et al 2014), the limiting factor to the current abyssal overturning system might be still mixing. Other simulations also showed an increase in overturning circulation with diffusivity (e.g., Toggweiler and Samuels 1998; Nikurashin and Vallis 2011).

3) *Their question 6: "Would the general circulation of the ocean be qualitatively different if Earth had no Moon?"* (Intriguing but without offering an answer in their paper)

   According to this study, mixing plays a dual role in the vigor of abyssal upwelling. Intensified mixing in the basin interior does expedite the erosion process and enhance abyssal upwelling. However, in the bottom-water formation region, heightened mixing tends to decrease the density of the bottom water, resulting in a weakening of the intrusion process and subsequently the upwelling (as seen in Case 4-SOe3 in Section 3). The ultimate outcome hinges on whether erosion or intrusion is more affected by the reduced mixing if Earth had no Moon. However, numerical studies suggest that a slowdown is the more likely outcome. Saenko and Merryfield (2005) discovered an "essentially stagnant deep North Pacific" if without tidal mixing. Additionally, Case 1 in Section 3 also observes a significantly diminished abyssal MOC with lower diffusivity. Regardless of the final conclusion, the global stratification structure, at least the abyssal stratification, would undergo a fundamental shift as it adjusts to a new equilibrium state without significantly reduced mixing.

An intriguing final question for discussion is what changes would take place within the ocean when the abyss transitions to a warming phase in the future. As mentioned in Section 2, the ongoing cooling trend in the abyssal ocean will cease over the next few hundred to thousand years. Instead, the warming signal that originated from the Antarctic atmosphere centuries ago, following the peak of the Little Ice Age, has likely already covered the Southern Ocean region and is anticipated to continue its path into the three tropical oceans, carried by AABW.

In line with the warming trend, the bottom-intensified dissipation would need to reverse its direction to promote convergence, rather than divergence, of vertical turbulent heat flux in the bottom ocean. Consequently, the diapycnal overturning circulation in Fig. 2d will also reverse, shifting to an anti-clockwise direction due to the positive $w_{dia}$ in response to an opposite flux pattern as per Eq. (5). In contrast, the Eulerian velocity, $w_{Eul}$, remains largely unaffected because the bottom-water erosion is not contingent on the vertical structure of turbulent fluxes.

As a result, the isopycnal displacement would lag behind the material surface during upwelling owing to the positive diapycnal velocity, $w_{dia} = w_{Eul} - w_{iso} > 0$. The isopycnal would either still rise, but at a slower rate than the material surface, or it could even descend.

Another question arises: how could the turbulent dissipation (or equivalently, the buoyancy flux), which is currently bottom-intensified, transition to a state of bottom-attenuation subject to a warming abyss? If we decompose the diffusion term in Eq. (4),

$$\frac{\partial}{\partial z}\left(\kappa \frac{\partial \rho}{\partial z}\right) = \frac{\partial \kappa}{\partial z}\frac{\partial \rho}{\partial z} + \kappa \frac{\partial^2 \rho}{\partial z^2}, \qquad (10)$$

the last two terms in RHS are opposite in sign given a bottom-intensified diffusivity. Current observations suggest that "the turbulent density flux is dominated by the decrease in diffusivity away from the ocean bottom, $(\partial_z \kappa)(\partial_z \rho) > 0$, and not by the increase in stratification, $\kappa \partial_z^2 \rho < 0$" (Ferrari et al 2016). This leads to a positive diffusion term in the LHS. Nevertheless, for a warming scenario, it is possible that the stratification transitions to a state that the relative magnitudes of the two terms overturn, leading to a negative density diffusion term. That is, in the context of abyssal warming or regions where warming is already underway, we may continue to observe a bottom-intensified diffusivity ($\kappa$), but hardly a bottom-intensified dissipation rate ($\varepsilon$). Another possibility for $\kappa$ in the warming phase is to simply become bottom-attenuated or uniform.

Lastly, Munk's theory is well applicable in the upper ocean (Wunsch 2023), which could be attributed to the relatively weak long-term trend in isopycnal displacement in that region (above 2000 m in Fig. 2c). Here, for the long-term average, $w_{iso}$ is approximately equal to $w_{Eul}$, and thus the steady-state assumption holds fine. However, caution should be exercised when applying the recipes in circumstances where isosurfaces displace at a rate comparable to the Eulerian velocity. For density (isopycnal), such circumstances are characterized by the dominance of adiabaticity, such as processes in the deep ocean or at shorter timescales (e.g., Han 2021,2023).

The erosion-intrusion mechanism does not refute the upwelling BBL theory. It serves as a complementary explanation for the cases of upwelling which BBL theory does not apply. This study does not provide evidence to dismiss the possibility of abyssal water being upwelled along the slope, as predicted by the BBL theory.

Some results from the numerical experiments remain unclear. For instance, why does the intrusion of bottom water tend to intensify to the west? Why does the intrusion of bottom water experience a slight disruption when approaching the Equator, even for a flat bottom (as indicated by the MOC streamfunctions)? Why does the wind forcing have a minimal influence on the abyssal MOC? Further explorations are solicited.

*Acknowledgments*.

I appreciate Weidong ZHAI and Chenglong LI for their patient discussions on ocean tracers, and Qiang DENG for assisting with running MITgcm. This work was supported by grant XMUMRF/2022-C9/ICAM/0009 from the Xiamen University Malaysia Research Fund.

*Data Available statement*.

The ECCO reanalysis dataset, version 4 release 3, is available at the website ecco.jpl.nasa.gov for open access. GLODAPv2 data used in this study was accessed from: https://glodap.info/index.php/merged-and-adjusted-data-product-v2-2022/ .

# Navigating Munk's Abyssal Recipes: Reconciling the Paradoxes and Suggesting a Mechanism for Bottom Water Upwelling in a Flat-Bottom Ocean

Lei Han [a, b]

[a] *China-ASEAN College of Marine Sciences, Xiamen University Malaysia, Sepang, Malaysia*

[b] *College of Ocean and Earth Sciences, Xiamen University, Xiamen, China*

*Corresponding author*: Lei Han, lei.han@xmu.edu.my

## Supplementary materials

This document includes
- One supplementary table, Table S1;
- Five supplementary figures, from Fig. S1 to Fig. S5;

| Parameters (variable name in the MITgcm) | Values |
| --- | --- |
| Coordinate | Cartesian |
| Domain | 60°S-60°N, 0-130°E |
| Horizontal resolution (delX, delY) | 2°×2° |
| Vertical resolution (delR) | 80 m |
| Basin depth | 4000 m |
| SST restoring profile and timescale | refer to FMM16 |
| Wind stress | refer to FMM16 |
| Time step (deltaT) | 3000 s |
| EOS type (eosType) | Linear |
| Thermal expansion coefficient $\alpha$ (tAlpha) | $2\times10^{-4}$ °C$^{-1}$ |
| haline contraction coefficient $\beta$ (sBeta) | 0 |
| GM package | enabled |
| Horizontal viscosity (viscAh) | $2\times10^{5}$ m²/s |
| Vertical viscosity (viscAr) | $1.2\times10^{-4}$ m²/s |
| Bottom drag coefficient (bottomDragLinear) | $5\times10^{-4}$ m/s |
| Coriolis parameter at southernmost point (f0) | $-1.26\times10^{-4}$ s$^{-1}$ |
| Beta-plane parameter (beta) | $2.1\times10^{-11}$ m$^{-1}$s$^{-1}$ |

Table S1. Model parameters of the control run. All the parameters are referred to the FMM16 configuration.

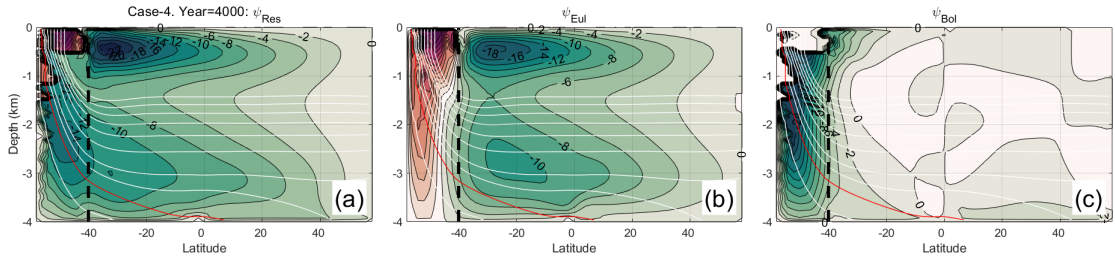

Fig. S1. The residual (a), Eulerian (b), and eddy-induced (c) MOC streamfunctions for the control run of MITgcm. The vertical dashed line indicates the northern boundary of the periodic channel. Temperature contours are indicated by the white lines, with the red line highlighting the 3°C one.

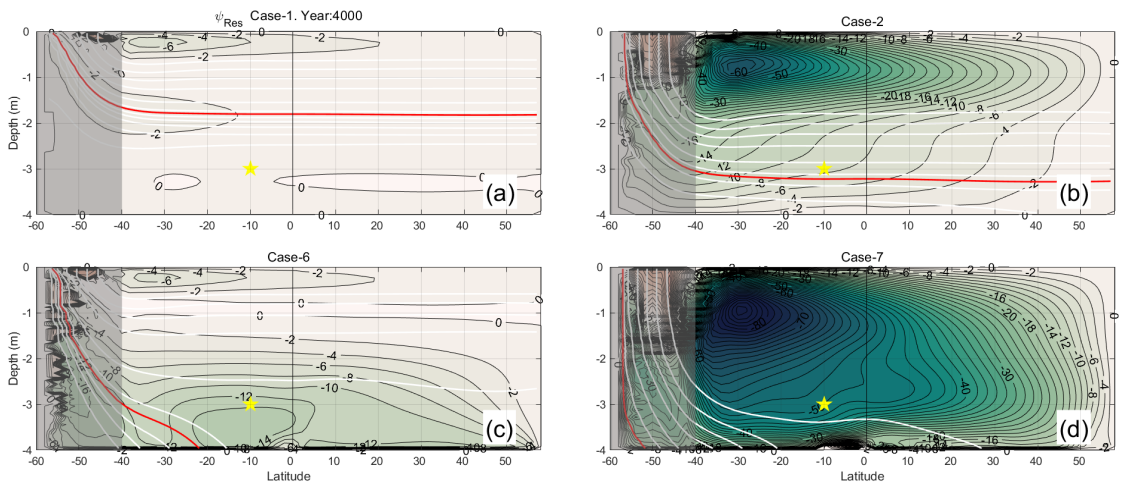

Fig. S2. Same as Fig. 4, but for Cases 1, 2, 6, and 7.

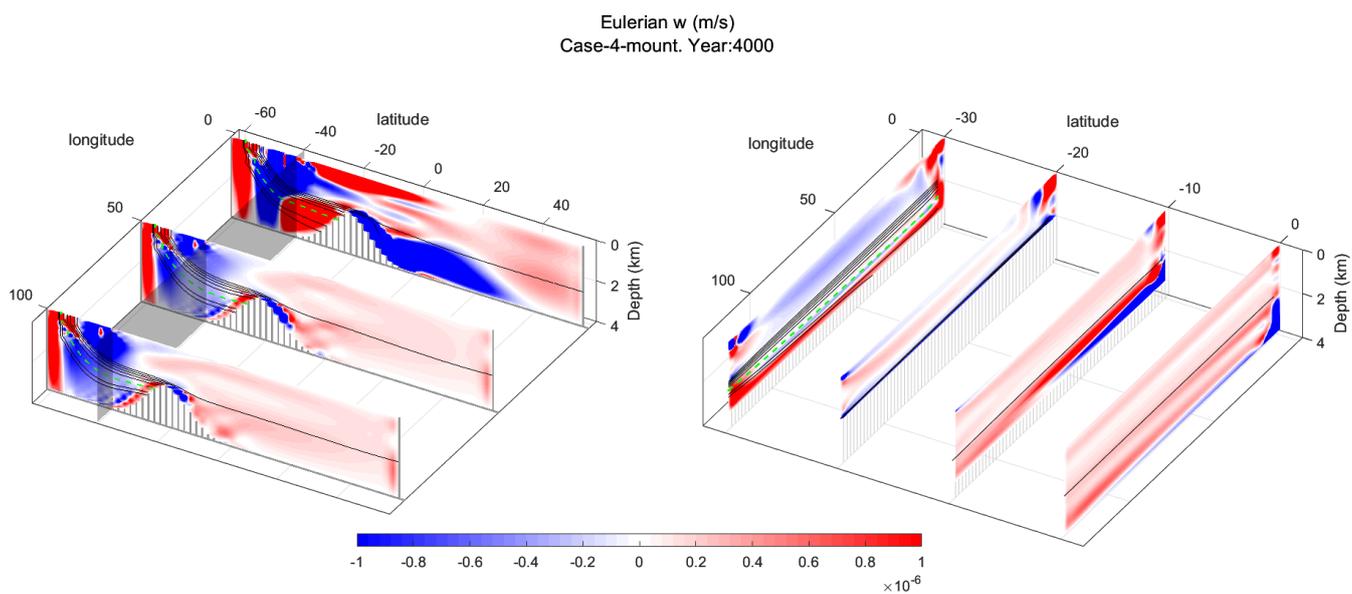

Fig. S3. Same as Fig. 5, but for Cases 4-mount and zonal sections of different latitudes shown (right panel).

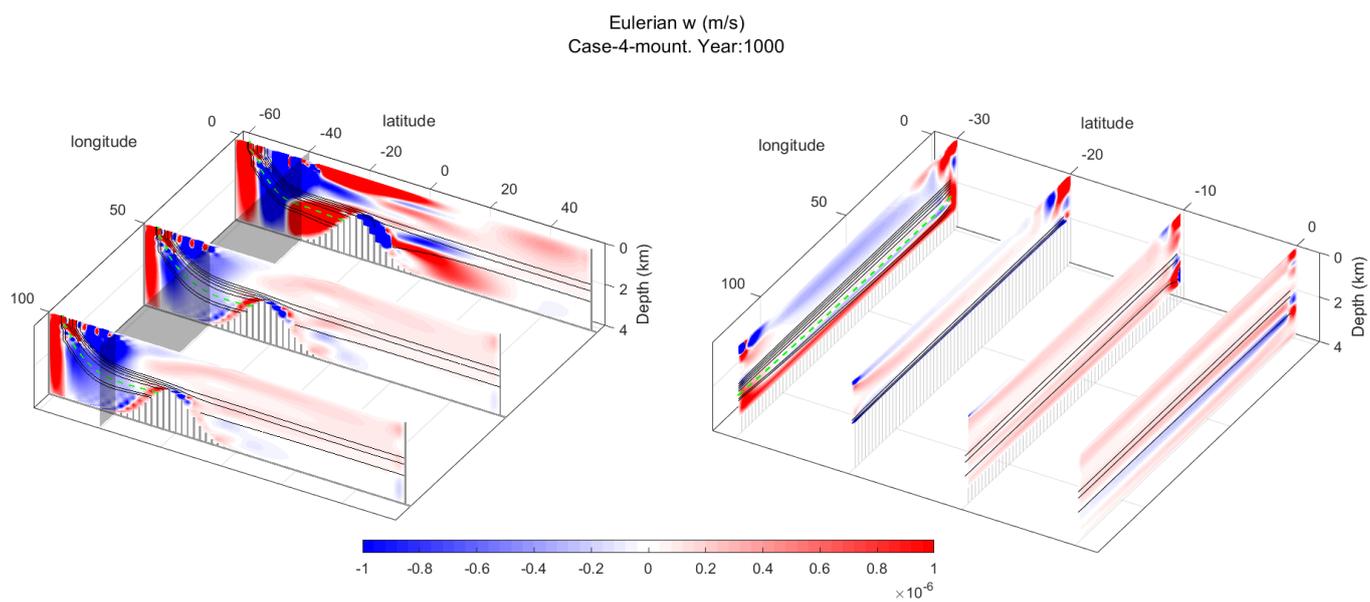

Fig. S4. Same as Fig. S3, but for Year 1000.

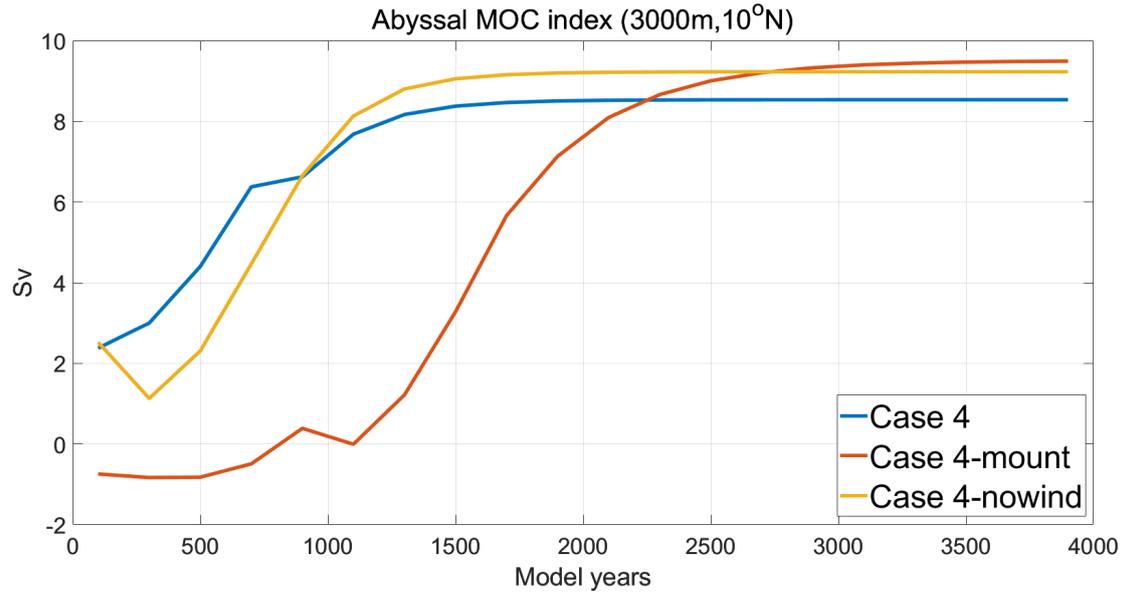

Fig. S5. Time series of the abyssal MOC index at 3000 m and 10°N for three cases: Case 4: control run; Case 4-mount: control run, but with the seamount topography; Case 4-nowind: control run, but without wind forcing.